\documentclass{amsart}%
\usepackage{amssymb}
\usepackage{amsmath}
\usepackage{amsfonts}%
\usepackage{graphicx}
\providecommand{\U}[1]{\protect\rule{.1in}{.1in}}

\theoremstyle{plain}

\numberwithin{equation}{section}
\begin{document}
\title[Emergence of Function]{Emergence of Function \ }
\author{Indika Rajapakse}
\address{\noindent University of Michigan}
\email{indikar@umich.edu}
\author{Steve Smale}
\address{University of California Berkeley}
\email{smale@math.berkeley.edu}
\thanks{We extend thanks to James Gimlett and Srikanta Kumar at Defense Advanced
Research Projects Agency for support and encouragement.\bigskip\ }

\begin{abstract}
This work gives a mathematical study of tissue dynamics. We combine
within-cell genome dynamics and diffusion between cells, where the synthesis
of the two gives rise to the emergence of function. We introduce a concept of
monotonicity and prove that monotonicity together with hardwiring, defined as
all cells of the same tissue having the same genome dynamics, is sufficient
for the global convergence of the tissue dynamics.

\end{abstract}
\maketitle

\noindent One of the most beautiful questions in biology is how individual
cells and tissues, each expressing information from a single genome, give rise
to all functions in a multicellular organism. Is there a basis for the
emergence of tissue-specific function? In vertebrates, consider the liver,
functioning to detoxify and ensure an appropriate composition of blood, and
the skeletal muscle, functioning to contract and generate\textbf{\ }%
force\textbf{.} In each of these tissues millions of individual cells
contribute to emergence of function according to their cell type.\ \smallskip

\noindent Tissue here means a set of cells of the same cell type located
together as a exemplified by an organ in the body.\smallskip

\noindent Understanding proteins is central to understanding this emergence
from single cells to a whole tissues. The main elements of emergence that we
consider are first, the unique protein distribution in a given cell type and
second, the cellular architecture of the tissue, a three dimensional structure
with diffusion of molecules between cells. \smallskip

\noindent We build a mathematical model for emergence of function, drawing on
our previous work on cell dynamics (genome dynamics), and the work of Alan
Turing on diffusion [1]. Here we combine the cell dynamics and the diffusion
between cells, where the synthesis of the two gives rise to the emergence of
function. Conditions are investigated under which the dynamics of the tissue
of an organism converge to an equilibrium where the proteins of individual
cells have the same distribution. Underlying our setting are known biological
phenomena: 1) cells within a tissue (i.e. the same cell type) have the same
dynamics and distribution of proteins at equilibrium 2) the function of a cell
corresponds to the proteins of that cell. For reasons to be discussed, we call
the property in 1) "hardwiring" of the tissue [2]. Convergence of the tissue
dynamics to such an equilibrium naturally takes on importance, for its role in
maintenance of tissue function. Even a local stability of the (hardwiring)
equilibrium, i.e. it's robustness, gives some validity to our model in
biology. We introduce a property of cell dynamics (first for one cell and then
extended to many cells) that we call monotonicity. Our main theorem (Theorem
5) establishes that monotonicity implies global convergence of the tissue
dynamics to the equilibrium, where all cells have the same protein
distribution. This gives a biological justification for our framework, and a
model for "emergence of function," as well as suggestions for studying the
passage from emergence to morphogenesis.\medskip

\noindent We cover the following content in this work: \medskip

\noindent1. A simple example

\noindent2. One cell and one protein from the gradient point of view

\noindent3. The genome dynamics of one cell and $n$ proteins

\noindent4. Cellular dynamics with a single protein

\noindent5. Dynamics of a tissue ($m$ cells and $n$ proteins)

\noindent6. Turing's paper on morphogenesis

\noindent7. Lapse of emergence\bigskip

\noindent\textbf{1. Simple example\medskip}

\noindent Here we model two cells, separated by a membrane, that each have\ a
same single protein. Consider the following system,%
\begin{align}
\text{Cell 1}\text{: \ \ \ \ }  & \frac{dx}{dt}=a\left(  x-x_{0}\right)
\tag{1a}\\
\text{Cell 2}\text{: \ \ \ \ }  & \frac{dy}{dt}=b\left(  y-y_{0}\right)
,\text{ \ }a,b,<0,x_{0},y_{0}>0,\tag{1b}%
\end{align}

\noindent where $x$ and $y$ can be interpreted as\ protein concentration, $x$
in cell 1 and $y\ $cell 2, both positive. Thus, \ $x,y\in\mathbf{X\times Y=} $
$\left[  0,c\right]  \mathbf{\times}\left[  0,d\right]  ,$ where $c$ and $d$
represent the maximum concentration of protein $x$ and $y$
respectively.\noindent\ The equilibria are$:x=x_{0}$ and $y=y_{0}.$ \medskip

\noindent We introduce Turing type (diffusion) coupling by adding a term with
$\beta>0$ as follows:%

\begin{align}
\frac{dx}{dt}  & =a\left(  x-x_{0}\right)  +\beta(y-x)\tag{2}\\
\frac{dy}{dt}  & =b\left(  y-y_{0}\right)  +\beta(x-y),\text{ \ }%
\beta>0\nonumber
\end{align}

\noindent The equilibrium for the above system is obtained by solving the
system derived from Equation 2) by setting the right hand sides equal to zero.
This is a linear system in two equations and two variables and we obtain
\begin{align}
x  & =\frac{-abx_{0}+a\beta x_{0}+b\beta y_{0}}{a\beta+b\beta-ab}\tag{3a}\\
y  & =\frac{-aby_{0}+a\beta x_{0}+b\beta y_{0}}{a\beta+b\beta-ab}.\tag{3b}%
\end{align}

\noindent It is not hard to see from Equation\noindent\ 3a and 3b that if
$\beta\longrightarrow\infty$, $x$ and $y$ converge to the same value
$\frac{ax_{0}+by_{0}}{a+b}.$ \ Therefore this system approaches a common
protein concentration and the example exhibits the role of diffusion, even
with different cell dynamics.\ We refer to this as an "emergent equilibrium."
\ Eigenvalues of the Jacobian matrix (see the following sections) for system 2
at equilibrium and finite $\beta$ are expressed as:
\begin{align*}
& \frac{1}{2}\left(  a+b\right)  -\beta+\frac{1}{2}\sqrt{\left(  a-b\right)
^{2}+4\beta^{2}}\\
& \frac{1}{2}\left(  a+b\right)  -\beta-\frac{1}{2}\sqrt{\left(  a-b\right)
^{2}+4\beta^{2}}.
\end{align*}
\noindent Since the eigenvalues are real negative, this pair $\left(
x,y\right)  $ of Equation 3 is a stable equilibrium.\bigskip

\noindent\textbf{Success of emergence}: The magnitude of $\left(  x-y\right)
$ from Equation 3 measures the departure from the "emergence" as,
\begin{equation}
x-y=-\frac{ab\left(  x_{0}-y_{0}\right)  }{\beta\left(  a+b\right)
-ab}.\tag{4}%
\end{equation}
If $x_{0}-y_{0}$ is big and $\beta$ is small there is ill-conditioning as
follows. If $a=0$ in Equation (3), the solution is $x=$\ $y_{0}$ and
$y=y_{0}.$ If $a\neq0$ no matter how small$,$ and $\beta=0$, the solution is
$x=x_{0}$ and $y=y_{0}$\textbf{\medskip\ }

\noindent Note from Equation 4, for any finite $\beta$ the equilibrium for the
pair $\left(  x,y\right)  $ has $x$ not $y$ if $x_{0}\neq y_{0}$. We might say
then the system 2 is not emergent (for any finite $\beta$).\medskip

\noindent Figure 1 show a numerical example of this system.\medskip%
\begin{figure}[ptb]%
\centering
\includegraphics[
natheight=5.000400in,
natwidth=7.073300in,
height=2.028in,
width=2.8565in
]%
{../../Users/indikar/Desktop/em2.tif}%
\caption{Plot of Equations 3a and 3b for\ $a=-2,b=-1,c=-1,d=-2$ as $\beta$
changes. $x$ (red) and $y$ (blue) are the coordinates of the equilibrium of
Equation 2.}%
\end{figure}

\noindent\textbf{Remark 1}: Here the $\beta$ anticipates the fiedler number of
a laplacian defined by the cellular network of the tissue. \noindent We will
introduce the concept of a "Hardwiring hypothesis", which implies $x_{0}%
=y_{0}$. \ Diffusion is unnecessary for emergence (and in fact it can defeat
emergence (!) as we will see). On the other hand diffusion can have a
stabilizing effect.\bigskip

\noindent\textbf{2. One cell and one protein from the gradient point of view
\medskip}

\noindent We given an alternative point of view of genome dynamics that will
not be used in the rest of the paper. This section may therefore be safely
skipped.\ Consider the dynamics described as the gradient of a potential
function. \textbf{\medskip}

\noindent Suppose $f$ \ is quadratic\textbf{\ }%
\begin{equation}
f\left(  x\right)  =\frac{a}{2}\left(  x-x_{0}\right)  ^{2}+b,\text{
\ \ \ \ }x>0,x_{0}>0,a,b>0\tag{5}%
\end{equation}
where the derivative $f^{^{\prime}}\left(  x\right)  =a\left(  x-x_{0}\right)
,\arg\min f\left(  x\right)  =x_{0},$ $f\left(  x_{0}\right)  =b$ and $a$ is
the rate of convergence to the equilibrium (see Figure 2). $\ f$ could be
interpreted as a potential function and $x$ as a protein concentration. From
Equation 5, we obtain an example of genome dynamics as is in Section 1.%

\begin{equation}
\frac{dx}{dt}=-\left(  \text{Gradient}f\left(  x\right)  \right)  ,\text{
\ \ }x>0\tag{6}%
\end{equation}

\noindent\noindent where the derivative $f^{^{\prime}}\left(  x\right)
=a\left(  x-x_{0}\right)  ,$ $\min f\left(  x\right)  =b.$ Substituting
Equation 6 into Equation 5 we obtain an example of genome dynamics as is in
Section 1.%
\begin{equation}
\frac{dx}{dt}=-a\left(  x-x_{0}\right)  .\tag{7}%
\end{equation}
The solution to Equation 7 is%
\begin{equation}
x\left(  t\right)  =C\exp^{-at}+x_{0}\text{ \ \ \ \ \ }\tag{8}%
\end{equation}
checked as follows
\begin{equation}
\frac{dx\left(  t\right)  }{dt}=-aC\exp^{-at}=-a\left(  x\left(  t\right)
-x_{0}\right)  .\tag{9}%
\end{equation}
Solving for $C$ by setting $t=0$ in Equation 9%
\[
C=x\left(  0\right)  -x_{0},\ \ \ x\left(  0\right)  =\text{initial\ condition
of }(8)
\]
Therefore \ \ \ \textbf{\ }
\[
x\left(  t\right)  =\left(  x\left(  0\right)  -x_{0}\right)  e^{-at}%
-x_{0},\text{ \ \ }x_{0}>0,\text{ }a>0.
\]
The equilibrium for $\frac{dx}{dt}=-a\left(  x-x_{0}\right)  $ is given by
$x_{0}.$ Since the second derivative of $f$ in Equation 6 is negative, the
equilibrium is a stable equilibrium.\textbf{\medskip}

\noindent Figure 2 is an example of a potential function and its
$-$gradient.\textbf{\medskip}
\begin{figure}[ptb]%
\centering
\includegraphics[
natheight=4.260100in,
natwidth=5.176800in,
height=2.1577in,
width=2.6161in
]%
{../../Users/indikar/Desktop/port1.tif}%
\caption{The blue curve is the potential function $f$ $\left(  x\right)
=2\left(  x-x_{0}\right)  ^{2}+1$ and the black curve is the negative gradient
of $f\left(  x\right)  .$}%
\end{figure}
\ \ \ \ 

\noindent This is an example of linear dynamics of one protein and one cell
with stability. This dynamics while linear is also a good approximation of the
general stable dynamics in a neighborhood of the equilibria. \ Moreover the
global dynamics of the basin of the stable equilibrium is qualitatively
equivalent to the linear example. \bigskip\ 

\noindent\textbf{3. The genome dynamics of one cell and }$n$%
\textbf{\ proteins}\ \medskip

\noindent We use the setting of our paper on genome dynamics [2]. \ For a
single cell state space $\mathbf{X=}\underset{j}{\overset{n}{\prod}}$ $\left[
0,c_{j}\right]  ,$where $c_{j}$ is the maximum concentration of protein $j.$
Sometime we use $c_{j}=\infty.$ The inner product is Cartesian. The genome
dynamics is expressed as $\frac{d\mathbf{x}}{dt}=\mathbf{F}(\mathbf{x}),$
where $\mathbf{F}$ is a function from $\mathbf{X}\subset%
\mathbb{R}
^{n}$ to $%
\mathbb{R}
^{n}.\medskip$ \ 

\noindent Generally recall the notion of stable equilibrium $\mathbf{x}_{0}$
of $\mathbf{X}$ for $\frac{d\mathbf{x}}{dt}=\mathbf{F}(\mathbf{x})$, as well
as its basin. \ If every eigenvalue of the Jacobian matrix of first partial
derivatives of $\mathbf{F}$ at \ $\mathbf{x}_{0}$ has negative real part, then
all trajectories that start near $\mathbf{x}_{0}$ approach it as
$t\rightarrow\infty.$\ The basin $\mathbf{B}\left(  \mathbf{x}_{0}\right)
$\textit{\ }is the set of all points which tends to $\mathbf{x}_{0}$ when
$t\rightarrow\infty.$ Then $\mathbf{x}_{0}$ is a stable equilibrium.\medskip

\noindent The dynamics on the basin is "linear" provided that the $\mathbf{F}
$ $:\mathbf{B}\rightarrow%
\mathbb{R}
^{n}$ $,$ $\mathbf{B}=%
\mathbb{R}
^{n},$ \ has the form $\mathbf{F}\left(  \mathbf{x}\right)  =\mathbf{F}%
_{lin}\left(  \mathbf{x-x}_{0}\right)  $, where $\mathbf{F}_{lin}$ is a linear
map, $%
\mathbb{R}
^{n}\rightarrow%
\mathbb{R}
^{n},$ $\mathbf{x}$ belongs to $\mathbf{B}$ and $\mathbf{x}_{0}$ is the
equilibrium in $\mathbf{B}.\medskip$

\noindent Suppose that $\mathbf{F}$ is the dynamics, not necessarily linear,
with stable equilibrium $\mathbf{x}_{0}$. Then $\mathbf{F}$ at $\mathbf{x}%
_{0}$ has the above form in a neighborhood of $\mathbf{x}_{0}$ ($\mathbf{F}%
_{lin}$ is the derivative of $\mathbf{F}$ at $\mathbf{x}_{0}).$ More over it
can be shown that the dynamics on the basin is topologically equivalent to the
linear dynamics as above.\medskip

\noindent In fact a main theorem about stable equilibria is that the linear
dynamic in $%
\mathbb{R}
^{n}$\ is equivalent to the dynamics of $\frac{d\mathbf{x}}{dt}=\mathbf{F}%
\left(  \mathbf{x}\right)  $\ in the basin $\mathbf{B}\left(  \mathbf{x}%
_{0}\right)  .$\ This means there is a homeomorphism from $%
\mathbb{R}
^{n}$\ to the basin $\mathbf{B}\left(  \mathbf{x}_{0}\right)  $\ that
preserves the solution curves. \ Then the dynamics of $\mathbf{B}\left(
\mathbf{x}_{0}\right)  $\ is same topologically as linear dynamics above.
\medskip

\noindent Browder [3, 4] and Hirsch [5, 6] have extensively studied the topic
of monotonicity. Here we provide our version of monotonicity which is new, has
a common element with Hirsch and Browder, but is quite different.\medskip

\noindent\textbf{Definition 1: Monotonicity condition}. Suppose we have a
dynamics $\frac{d\mathbf{x}}{dt}=\mathbf{G}(\mathbf{x})$ on a domain
$\mathbf{X}$ in a euclidean space with it's inner product. The monotonicity
condition for $\mathbf{G,}$ and a point $\mathbf{x}_{0}$ $\in$ $\mathbf{X}$ is
then:
\begin{equation}
\left\langle \mathbf{Gx,}\left(  \mathbf{x-x}_{0}\right)  \right\rangle
<0\text{ for all }\mathbf{x\neq x}_{0}\text{ in }\mathbf{X}.\tag{10}%
\end{equation}
$\medskip$

\noindent\textbf{Proposition 1:} For any dynamics $\frac{d\mathbf{x}}%
{dt}=\mathbf{G}(\mathbf{x})$ on $\mathbf{X\subset}$ $%
\mathbb{R}
^{k}\mathbf{,}$ $\mathbf{x}_{0}$ $\in$ $\mathbf{X},$ the monotonicity
condition for $\mathbf{X},$ $\mathbf{x}_{0},$ implies that $\left\Vert
\mathbf{x}\left(  t\right)  -\mathbf{x}_{0}\right\Vert $ is a decreasing
function of $t$ for all non trivial solutions $\mathbf{x}(t)$ in $\mathbf{X,}
$ where $\mathbf{x}(t)$ defined for all $t>0.$\medskip

\noindent\textbf{Proof:} \ Suppose Equation 10 is true. Note that
\begin{align*}
\frac{d\left(  \left\Vert \mathbf{x}\left(  t\right)  -\mathbf{x}%
_{0}\right\Vert ^{2}\right)  }{dt}  & =\frac{d\left\langle \mathbf{x}\left(
t\right)  -\mathbf{x}_{0},\mathbf{x}\left(  t\right)  -\mathbf{x}%
_{0}\right\rangle }{dt}\\
& =2\left\langle \frac{d\left(  \mathbf{x}\left(  t\right)  \right)  }%
{dt},\mathbf{x}\left(  t\right)  -\mathbf{x}_{0}\right\rangle =2\left\langle
\mathbf{Gx}\left(  t\right)  ,\left(  \mathbf{x}\left(  t\right)
-\mathbf{x}_{0}\right)  \right\rangle
\end{align*}

\noindent The quantity at the end is negative by Equation 10, the monotonicity
condition. QED\medskip

\noindent One could call the $\mathbf{X}$ of Proposition 1, a "monotonic
basin" for the dynamics. Under these conditions $\mathbf{x}_{0}$ is an
equilibrium.\medskip

\noindent Thus, monotonicity on $\mathbf{X,x}_{0}$ implies that $\mathbf{x}%
\left(  t\right)  $ is monotonically converging to $\mathbf{x}_{0}.$ This
gives a stability of $\mathbf{x}_{0}$. The converse is not true not even in
the linear case. One can take for an example a spiral sink where the axes are
different (Figure 3). When the solution is going in the direction of the long
axis then $\mathbf{x}\left(  t\right)  -\mathbf{x}_{0}$ is not decreasing,
while $\mathbf{x}_{0}$ is a stable equilibrium. This example helps understand
the famous Turing phenomenon (see Section 6).
\begin{figure}[ptb]%
\centering
\includegraphics[
natheight=2.302100in,
natwidth=3.239600in,
height=1.4088in,
width=1.9726in
]%
{../../Users/indikar/Desktop/mon1.tif}%
\caption{$\left\langle \mathbf{F}p,\left(  p-\mathbf{x}_{0}\right)
\right\rangle >0.$ This an example of a basin which is not a monotonic basin.}%
\end{figure}

\medskip\noindent\textbf{Example 1}: \noindent Let $\mathbf{Fx=A}\left(
\mathbf{x-x}_{0}\right)  ,$ where $\mathbf{A}$ is a linear map $%
\mathbb{R}
^{n}\rightarrow%
\mathbb{R}
^{n},$ not necessarily symmetric. Then $\mathbf{A}$ is negative definite
exactly when monotonicity holds.\bigskip

\noindent Let us return to the biological setting. Single cell dynamics is
that of dynamics on a basin $\mathbf{B}\subset\mathbf{X}$ as in our previous
work on genome dynamics \textbf{[}2\textbf{]}. We assume that the basin
$\mathbf{B}$ is that of an equilibrium $\mathbf{x}_{0}$ and are excluding
periodic attractors in the present paper. This means we are identifying a cell
with its basin. The equilibrium of a genome dynamics of a cell exhibits the
distribution of proteins. That distribution can be identified with that
cell.\medskip

\noindent We now examine explicitly the conditions for monotonicity in the
linear case of one cell with two proteins. This case can be represented by the
following system represented by the following system%
\begin{equation}
\frac{d\mathbf{x}}{dt}=\mathbf{Fx=A}\left(  \mathbf{x}-\mathbf{x}_{0}\right)
,\text{ \ }\mathbf{x=}\text{ }\left(  \mathbf{x}_{1},\mathbf{x}_{2}\right)
\nonumber
\end{equation}

\noindent where $\mathbf{A}=\left(
\begin{array}
[c]{cc}%
a & b\\
c & d
\end{array}
\right)  $ and $\frac{d\mathbf{x}}{dt}=0$ when $\mathbf{x}=\mathbf{x}_{0}.$
Then $\mathbf{A}$ (Jacobian matrix at $\mathbf{x}_{0}$) is stable and
$\mathbf{x}\left(  t\right)  \rightarrow\mathbf{x}_{0}$ when\ all eigenvalues
have negative real parts. $\ $The eigenvalues of $\mathbf{A}$ are given by the
characteristic equation $\lambda^{2}-\tau\lambda+\Delta=0,$ where
\[
\tau=\text{trace}(\mathbf{A)=}\text{ }a+d\text{ and \ }\Delta=\text{det}%
(\mathbf{A)=}\text{ }ad-bc.
\]
Then $\lambda_{1}=\frac{\tau+\sqrt{\tau^{2}-4\Delta}}{2},\lambda_{2}%
=\frac{\tau-\sqrt{\tau^{2}-4\Delta}}{2}$ are the eigenvalues of $\mathbf{A.}$
For stability $\mathbf{A}$ must satisfy two criteria: 1) The trace, $a+d,$
must be negative, and 2) the determinant, $ad-bc,$ must be positive [6, 7].
\ \medskip

\noindent To derive the conditions for monotonicity, consider the quadratic
form associated with $\mathbf{A:}$ $Q(u,v)=au^{2}+dv^{2}-\left(  \frac{b+c}%
{2}\right)  uv,$ $\ \ $and suppose $a,d<0.$ Thus, $\frac{\mathbf{A+A}^{T}}%
{2}=\left(
\begin{array}
[c]{cc}%
a & \frac{b+c}{2}\\
\frac{b+c}{2} & d
\end{array}
\right)  ,$ and $\frac{\mathbf{A+A}^{T}}{2}$ is symmetric. The matrix of a
quadratic form can always be forced to be symmetric in this way. \ The
condition for monotonicity is $\left\langle \mathbf{Fx},\left(  \mathbf{x}%
-\mathbf{x}_{0}\right)  \right\rangle <0.$ This amounts to $\left\langle
\mathbf{A}\left(  \mathbf{x}-\mathbf{x}_{0}\right)  ,\left(  \mathbf{x}%
-\mathbf{x}_{0}\right)  \right\rangle <0$ or that the eigenvalues of $\left(
\frac{\mathbf{A+A}^{T}}{2}\right)  $ are negative$,$ which is equivalent to
$\mathbf{A}$ being negative definite. Since the determinant of $\frac
{\mathbf{A+A}^{T}}{2}$ is positive,
\begin{equation}
ad>\left(  \frac{b+c}{2}\right)  ^{2},\text{ \ }a,d<0.\tag{11}%
\end{equation}
In summary the stability condition is $bc<ad$ \ and the monotonicity condition
is $\left(  \frac{\left(  b+c\right)  }{2}\right)  ^{2}<ad.$ Therefore, excess
of the left hand sides of the previous inequalities is $\left(  \frac{\left(
b+c\right)  }{2}\right)  ^{2}-bc.$ If the excess is positive or zero,
monotonicity implies stability. The excess is never negative. \medskip

\noindent More generally, as a consequence of Proposition 1 one can prove the
following.\medskip

\noindent\textbf{Proposition 2}: For a linear dynamics on $%
\mathbb{R}
^{n}$ monotonicity implies stability.\medskip

\noindent Figure 4 shows an example of monotonicity and stability conditions
in the $bc$ plane, where $a,d=-1$. $\mathbf{E}$ is the monotonic region and
hence is part of the stable region. The red dot represents Turing's two-cell
example (Turing [1] and Chua [11]), discussed in Section 6.\medskip\
\begin{figure}[ptb]%
\centering
\includegraphics[
natheight=4.337000in,
natwidth=5.631700in,
height=2.1958in,
width=2.8435in
]%
{../../Users/indikar/Desktop/mon2.tif}%
\caption{Monotonicity and stability conditions in the $bc$ plane for $a,d=-1$.
\ The dark shading region together with the blue region (\textbf{E})
constitutes the stability region. \textbf{E }is the monotonic region.\ The red
dot shows Turing's two-cell example in the $bc$ plane. \ }%
\end{figure}

\noindent\textbf{Hardwiring:} The genes present in the human genome are the
same in all cell types and all individuals. Now we describe a property of a
family of cells, which we called hardwiring [2], motivated by the universality
above. Our network in [2] puts an oriented edge (between two nodes), between
two genes, $i$ and $j$, if it is possible for the protein product of gene $i$
to bind to the promoter of gene $j$ and activate transcription. Gene $i$ will
bind to this promoter only in some cell types, at certain stages of
development. It can happen that gene $i$ as a transcription factor may be
silenced. In that case gene $i$ can be removed from the network together with
its edges. As an example, this phenomenon can occur through failure of
chromatin accessibility [8, 9]. \ We will say that a family of cells is
hardwired\textbf{\ }provided that the genome dynamics is the same for every
cell in the family\textbf{. }In the example of Turing (also Smale [10], Chua
[11, 12]) below hardwiring is assumed extensively.\textbf{\medskip}

\noindent\textbf{Definition of weak hardwiring: }Thus, the family is hardwired
provided that the dynamics of each cell in the family is the same; in
particular the equilibrium of each cell is the same. That is, the protein
distribution at the equilibrium of each cell is the same. If the last property
is true then we will say that the family satisfies "weak hardwiring." The idea
of the weak hardwiring concept is that in a single cell type all cells have
the same equilibrium distribution of proteins [2]. This helps justify the
identification of a tissue with its protein distribution \bigskip

\noindent\textbf{4. Cellular dynamics and its architecture \medskip}

\noindent We will define a graph $\mathbf{G}$ as a mathematical model for the
cellular structure for a single tissue. First consider the $m$\ cells of the
single tissue and a single\ protein\textbf{. }The main biological object is
the cellular architecture of a tissue which consists of $m$ cells in three
dimensions. The graph $\mathbf{G}$ consists of nodes corresponding to the
cells of the tissue. The weighted edges of the graph are associated with the
membranes between two cells and define the notion of adjacency. This adjacency
is represented by a number which represents the diffusion, between two cells
and it depends on the interactions at their cell membranes [13]. This number
could be interpreted as the product of the permeability and the area of the
membrane between cell $i$ and cell $j$, a quantity represented by a matrix
element $a_{ij}$. We write $\mathbf{A=}$ $\left(  a_{ij}\right)  .$ The matrix
$\mathbf{A}$ is an $m\times m$ symmetric matrix, the adjacency matrix of the
architecture. \ Note that $\mathbf{A}$ does not depend on the protein levels.
Thus, $\mathbf{G}$ is a weighted graph whose nodes are $i=1,...,m,$ and edges
$a_{ij}$. We assume that the graph is connected. What we have discussed here
is a network whose nodes are cells and it not to be confused with the genome
network in Section 3. This model applies more literally to diffusion in the
case of small molecules.\medskip\ 

\noindent\textbf{Definition of a state}: A state associated to the graph is a
set of protein levels ${\mathbf{x}}_{1},\ldots,{\mathbf{x}}_{m}$, where
${\mathbf{x}}_{i}$ is the level of a single protein in the $i^{th}$ cell. Thus
a state ${\mathbf{x}}$ is a function of nodes $i$ and with value at node $i$
written as ${\mathbf{x}}_{i}$. The states form a linear space $\mathbf{S}$,
and feasible states, the subspace of functions with non-negative values
$\mathbf{S}_{+}$. The function ${\mathbf{x\in}}$ $\mathbf{S}$ is harmonic
provided that ${\mathbf{x}}_{i}$ is a constant function of $i.$ By our
hypothesis that the graph $\mathbf{G}$ is connected, it follows that the space
of harmonic functions is one dimensional. \medskip

\noindent For $n$ proteins we generalize the notion of ${\mathbf{x}}_{i}$ and
$\mathbf{S}.$ Now ${\mathbf{x}}_{i}$ is a distribution of proteins in the
$i^{th}$ cell, i. e. ${\mathbf{x}}_{i}=\left(  {\mathbf{x}}_{i}^{1}%
,\ldots,{\mathbf{x}}_{i}^{n}\right)  .$ Note that this expression can be
thought as a function of $i$. \ We assume that the membrane structure of the
tissue affects all the proteins equally. (this is a strong idealization, but
it can be relaxed easily as in Turing's example [1] in Section 6). \medskip

\noindent\textbf{The Laplacian matrix: }Let $\mathbf{D}$ be the diagonal
matrix with the $i^{th}$ element of the diagonal defined by $\underset{j}{\sum
}a_{ij}.$ Then the Laplacian is given by
\[
\mathbf{L}=\mathbf{D}-\mathbf{A}.\noindent\noindent\noindent
\]

\noindent$\mathbf{L}$ is a $\left(  m\times m\right)  $ real symmetric matrix,
and together with the non-negativity of the weights, this implies that it is
positive semidefinite [14, 15]. It is an operator on $\mathbf{S}$. \medskip

\noindent The diffusion dynamics defined by the cellular architecture may be
written as follows:
\[
\frac{d{\mathbf{x}}_{i}}{dt}=-\underset{j\in m_{i}}{%
{\displaystyle\sum}
}a_{ij}\left(  {\mathbf{x}}_{i}-{\mathbf{x}}_{j}\right)  ,\text{
\ \ \ \ \ }i=1,...,m.
\]
or
\begin{equation}
\frac{d{\mathbf{x}}}{dt}=-{\mathbf{Lx.}}\tag{12}%
\end{equation}

\noindent Note that Equation (12) is a linear system of ordinary differential
equations.\medskip

\noindent\textbf{Remark 2}: Harmonic functions are exactly set of
${\mathbf{x}}$, such that ${\mathbf{Lx=0.\medskip}}$

\noindent Note that our definition applies not just to a single protein, but
to an $n-$tuple ${\mathbf{x}}_{i}\mathbf{\ }$belonging to ${\mathbf{%
\mathbb{R}
}}^{n}.$ The feasible ones to $\left(  {\mathbf{%
\mathbb{R}
}}_{+}\right)  ^{n}.\mathbf{\medskip}$

\noindent\textbf{Proposition 3}: The system is globally stable with
equilibrium set, the harmonic functions$.\medskip$

\noindent\textbf{Proof}: $\ \left\langle -{\mathbf{Lx,x}}\right\rangle \leq0$
and $\left\langle -{\mathbf{Lx,x}}\right\rangle =0$ iff ${\mathbf{x=}}$
constant, i.e. ${\mathbf{x}}_{1}={\mathbf{x}}_{2}=\cdots={\mathbf{x}}%
_{m}{\mathbf{.}}$ The solution to Equation (12) will be denoted by
${\mathbf{x(}}t{\mathbf{)}}$ with initial conditions ${\mathbf{x(}%
}0{\mathbf{)=C,}}$ ${\mathbf{C\in%
\mathbb{R}
}}^{m}$. Now the solution is%

\[
{\mathbf{x}}\left(  t\right)  =e^{-\mathbf{L}t}\mathbf{C}.
\]

\noindent Then $\frac{d}{dt}\left\langle {\mathbf{x}}\left(  t\right)
{\mathbf{,x}}\left(  t\right)  \right\rangle =2\left\langle -{\mathbf{Lx}%
}\left(  t\right)  ,{\mathbf{x}}\left(  t\right)  \right\rangle <0,$ unless
${\mathbf{x}}\left(  t\right)  $ satisfies ${\mathbf{x}}_{1}\left(  t\right)
={\mathbf{x}}_{2}\left(  t\right)  =\cdots={\mathbf{x}}_{m}\left(  t\right)
=$ constant$.$ Therefore, the solution converges to a harmonic function.
QED\medskip

\noindent For the $n$ protein case, the harmonic functions form an
$n-$dimensional space defined by ${\mathbf{x}}_{1}={\mathbf{x}}_{2}%
=...={\mathbf{x}}_{m}$, \ where ${\mathbf{x}}_{1}$ is an arbitrary element of
${\mathbf{%
\mathbb{R}
}}^{n}.$ \bigskip

\noindent\textbf{5. Dynamics of a tissue (}$m$\textbf{\ cells and }%
$n$\textbf{\ proteins})\textbf{\ \medskip}

\noindent We use both the notations $\mathbf{X}_{i}%
=\overset{n}{\underset{j}{\prod}}$ $\left[  0,c_{j}\right]  $ and the basin,
$\mathbf{B}_{i}\subset\mathbf{X}_{i}$ with its equilibrium $\mathbf{x}_{0,i}$
for the domain of the genome dynamics, where $c_{j}$ is the maximum protein
concentration protein $j$ and $n$ is number of proteins. $\mathbf{X}_{i}$ is
important for the lapse of emergence and dealing with different cell types
(different tissues) as in Section 7. $\mathbf{B}_{i}$ is suited for single
tissue theory as in the following. \medskip

\noindent\textbf{Genome dynamics for }$m$\textbf{\ cells:}\medskip

\noindent For a single cell say $i,$ $\mathbf{B}_{i}$ is the domain of the
dynamics. For each cell $i$, $\mathbf{F}_{i}:\mathbf{B}_{i}\rightarrow%
\mathbb{R}
^{n}$ represents the genome dynamics in cell $i$,%

\begin{equation}
\frac{d{\mathbf{x}}_{i}}{dt}=\mathbf{F}_{i}{\mathbf{x}}_{i},\text{
\ }{\mathbf{x}}_{i}\in\mathbf{B}_{i},\text{ }i=1,...m.\tag{13}%
\end{equation}

\noindent For the case of cells of a tissue \textbf{\ }$\mathbf{S}=$
$\underset{i}{\overset{m}{\prod}}\mathbf{B}_{i},i=1,...,m,$\textbf{\ }%
where\textbf{\ }$m$\textbf{\ }is the number of cells in the tissue and
$\mathbf{S} $ is the state space of Section 3 extended to $n$ proteins. We use
an inner product on $\mathbf{S}$ derived from the inner products on
$\mathbf{B}_{i}$ \medskip

\noindent Thus $\mathbf{B}_{i}$ corresponds to cell $i$ with stable
equilibrium $\mathbf{x}_{0,i}.$\medskip\ 

\noindent Now we take the product of the dynamics over all the cells at once
to get $\mathbf{F:S}{}\rightarrow%
\mathbb{R}
^{N}\left(  \text{or better }\left(
\mathbb{R}
^{n}\right)  ^{m}\text{ }\right)  ,$ where $\mathbf{F=}$ $\left(
\mathbf{F}_{1},\cdots,\mathbf{F}_{m}\right)  ,$ $N=nm$ and
\begin{equation}
\frac{d{\mathbf{x}}}{dt}=\mathbf{F}{\mathbf{x,}}\text{ \ \ }{\mathbf{x\in}%
}\text{ }\mathbf{S.}\tag{14}%
\end{equation}
\noindent Equation 14 is rephrasing Equation 13. This is the genome dynamics
of the tissue.\medskip

\noindent Thus this tissue has a genome dynamics and separates into a
individual cell dynamics $\mathbf{B}_{i}$ for cell $i.$ Let $\mathbf{x}_{0}$
be the point of $\mathbf{S}$ defined as $\mathbf{x}_{0}=$ $\mathbf{(}%
{\mathbf{x}}_{0,1},{\mathbf{x}}_{0,2},...,{\mathbf{x}}_{0,m}),$ where
$\mathbf{x}_{0,i}$ is the equilibrium in $\mathbf{B}_{i}.$ The weak hardwiring
hypothesis asserts that the $\mathbf{x}_{0,i}$ are all the same$.$ Then
$\mathbf{x}_{0}$ is the equilibrium for genome dynamics for the whole
tissue.\medskip

\noindent The rest of the paper we assume the weak hardwiring for the cells in
the tissue.\medskip

\noindent\textbf{Extension of monotonicity from the genome dynamics of a cell
to the genome dynamics of the tissue \medskip}

\noindent Extension of the definition of\textbf{\ }monotonicity to many cells
is given by:%

\begin{align*}
\left\langle \mathbf{Fx,}\left(  \mathbf{x-x}_{0}\right)  \right\rangle  &
<0,\text{ }\mathbf{F=}\prod\mathbf{F}_{i},\text{ \ }\mathbf{x\neq x}_{0}\\
\mathbf{x}  & \mathbf{=}\left(  \mathbf{x}_{1},\mathbf{x}_{2},...,\mathbf{x}%
_{m}\right)  \text{ and }\mathbf{x}_{1}\text{ }\mathbf{=(}{\mathbf{x}}_{1}%
^{1},{\mathbf{x}}_{1}^{2},...,{\mathbf{x}}_{1}^{n})\in\mathbf{X}_{1}\text{,
etc.}%
\end{align*}
Here ${\mathbf{x}}_{i}^{j}$ denotes the amount of $j^{th}$ protein in the
$i^{th}$ cell. $\ \frac{d\mathbf{x}}{dt}=\mathbf{Fx}$ is the dynamics on the
basin $\mathbf{B}=\prod\mathbf{B}_{i}\mathbf{,}$ $\ \mathbf{x}_{0}=\left(
\mathbf{x}_{0,1},\mathbf{x}_{0,2},...,\mathbf{x}_{0,m}\right)  $ and
$\mathbf{x}_{0}\in$ $\mathbf{B}.$ Observe from weak hardwiring $\mathbf{x}%
_{0,1}=\mathbf{x}_{0,2}=...=\mathbf{x}_{0,m}.$\medskip

\noindent\textbf{Example 2}: $\mathbf{Fx=A}\left(  \mathbf{x-x}_{0}\right)  ,$
where $\mathbf{A=}$ $\prod\mathbf{A}_{i},$ so that $\mathbf{A}$ is a
multi-linear map and each $\mathbf{A}_{i}:$ $%
\mathbb{R}
^{m}\rightarrow%
\mathbb{R}
^{m}$ is linear$.$ Then $\mathbf{A}_{i},$ for each $i$ is negative definite
exactly when this monotonicity holds. \medskip

\noindent However we are not assuming the linearity of the dynamics. One
cannot even get a good model of robust stability of equilibria in the linear
setting. One cannot model dynamics with two separate equilibria.\textbf{\ }%
\medskip

\noindent\textbf{Diffusion} \textbf{dynamics for }$n$\textbf{\ proteins:}
Recall section 4, the diffusion dynamics between cells in a tissue for a
protein distribution \
\begin{equation}
\frac{d{\mathbf{x}}}{dt}=\mathbf{-}{\mathbf{Lx}}\text{ or }\frac{d{\mathbf{x}%
}_{i}}{dt}=\mathbf{-}{\mathbf{Lx}}_{i}\text{ for all }i=1,..,m.\tag{15}%
\end{equation}

\noindent Here ${\mathbf{x=}}$ $\left(  {\mathbf{x}}_{1},...,{\mathbf{x}}%
_{m}\right)  ,$ ${\mathbf{x}}_{i}{\mathbf{\in}}%
\mathbb{R}
^{n}$ is an $n$-tuple of proteins or "a distribution of proteins." \medskip

\noindent\textbf{Dynamics of a tissue\medskip}

\noindent We will make the hypothesis that the cells described by
$\mathbf{F}_{i}$ $:$ $\mathbf{B}_{i}\rightarrow%
\mathbb{R}
^{m}$ have the same dynamics and the $\mathbf{B}_{i}$ are the same. This is
the hardwiring hypothesis. \ We also suppose that the basins $\mathbf{B}_{i}$
are convex. These two hypothesis are made so that the diffusion terms in
Equation 16 below make sense.\textbf{\medskip}

\noindent Following the spirit of Turing's paper, we may combine two dynamics
(genome dynamics within the cell and diffusion dynamics between cells) into a
system (16) that is the object of the study of this paper. \
\begin{equation}
\frac{d{\mathbf{x}}}{dt}=\mathbf{F}{\mathbf{x}}-{\mathbf{Lx,}}\text{
\ \ }{\mathbf{x\in}}\mathbf{B}=\overset{m}{\underset{i=1}{\prod}}%
\mathbf{B}_{i}\tag{16}%
\end{equation}

\noindent We emphasize that differential equation 16 is not necessarily linear
in contrast to Turing. \medskip\ 

\noindent The main Theorem of this paper is: \medskip

\noindent\textbf{Theorem 5: }The dynamical system $\frac{d{\mathbf{x}}}%
{dt}=\mathbf{F}{\mathbf{x}}-{\mathbf{Lx}}$ of a tissue (Equation 16) is
globally stable with equilibrium $\mathbf{x}_{0}$ provided the tissue is
hardwired, the basins $\mathbf{B}_{i}$ are convex and $\mathbf{F}$ satisfies
monotonicity. \medskip

\noindent\textbf{Lemma 1: }If\textbf{\ }$\mathbf{F}$ is monotone relative to
$\mathbf{B,x}_{0}$, \ then $\left(  \mathbf{F-\mathbf{L}}\right)  $ is
monotone.\medskip

\noindent\textbf{Proof of Lemma 1}. Lemma 1 is true if $\mathbf{F}$ is
monotone and if $\mathbf{-L}$ is monotone. \ First $\left\langle
-\mathbf{Lx,}\left(  \mathbf{x}-\mathbf{x}_{0}\right)  \right\rangle \leq0$ is
proved$.$ \ From weak hardwiring $\mathbf{\mathbf{x}_{0}}$ is harmonic and so
$\mathbf{L\mathbf{x}_{0}=0.}$ \ Moreover $\left\langle -\mathbf{Lx,x}%
\right\rangle \leq0$ for any $\mathbf{x}\in\mathbf{S,}$ because of
$-\mathbf{\mathbf{L}}$ is negative semi-definite. See Section 4.\ Therefore
$\left\langle -\mathbf{Lx,}\left(  \mathbf{x}-\mathbf{x}_{0}\right)
\right\rangle \leq0.$ Since $\mathbf{F}{\mathbf{x}}_{0}$ ${\mathbf{=}}$
$\ 0{\mathbf{,}}$ thus it remains only to prove $\left\langle \mathbf{Fx,}%
\left(  \mathbf{x}-\mathbf{x}_{0}\right)  \right\rangle <0$ for $\mathbf{x}%
\neq\mathbf{x}_{0}.$ \ But $\mathbf{F}$ is monotone by hypothesis. \ QED.
\medskip

\noindent By Proposition 1 applied to $\mathbf{G=}$ $\left(
\mathbf{F-\mathbf{L}}\right)  $ and Lemma 1 we obtain the global stability of
equilibrium ${\mathbf{x}}_{0},$ thus proving Theorem 5. \medskip

\noindent We will name the property of $\frac{d{\mathbf{x}}}{dt}%
=\mathbf{F}{\mathbf{x}}-{\mathbf{Lx}}$ in Theorem 5 "emergence."\medskip

\noindent Theorem 5 establishes that monotonicity implies global convergence
of the tissue dynamics to the equilibrium, that is all cells have the same
protein distribution, in a strong stable sense ("robustness"). This gives a
biological justification for the concept of hardwiring in a tissue. Thus we
give a model for "emergence of function."\bigskip

\noindent\textbf{Remark 3:} \textbf{Explicit solution of }$\frac{d{\mathbf{x}%
}}{dt}=\mathbf{F}{\mathbf{x}}-{\mathbf{Lx}}$ \textbf{in the linear
case\medskip}

\noindent Recall the linear case%
\begin{equation}
\frac{d{\mathbf{x}}}{dt}=\mathbf{F}\left(  {\mathbf{x-x}}_{0}\right)
-{\mathbf{Lx,}}\text{ \ \ \ \ }{\mathbf{x=}}\text{ }\left(  {\mathbf{x}}%
_{1}\mathbf{,...,}{\mathbf{x}}_{m}\right)  .\tag{17}%
\end{equation}
where $\left(  \mathbf{F}-{\mathbf{L}}\right)  $ is not singular. This is in
the form%

\begin{equation}
\frac{d{\mathbf{x}}}{dt}=\mathbf{P}{\mathbf{x}}-\mathbf{Q.}\tag{18}%
\end{equation}
where $\mathbf{P=F-L}$ and $\mathbf{Q=F}{\mathbf{x}}_{0}.$ Equation 18 has an
explicit solution [16]%
\[
{\mathbf{x}}(t)=\operatorname{exp}(\mathbf{P}t)\mathbf{C}+\mathbf{P}%
^{-1}\mathbf{Q.}%
\]
${\mathbf{x}}\left(  0\right)  =\mathbf{C+}$ $\mathbf{P}^{-1}\mathbf{Q,}$
therefore $\mathbf{C}=$ ${\mathbf{x}}\left(  0\right)  -\mathbf{P}%
^{-1}\mathbf{Q.}$ Furthermore, $\lim_{t\rightarrow\infty}{\mathbf{x}%
}(t)=\mathbf{P}^{-1}\mathbf{Q}$ if the eigenvalues of $\mathbf{P}$ strictly
negative.\bigskip

\noindent\textbf{Section 6:} \textbf{Turing's paper on morphogenesis\medskip}

\noindent The work of Alan Turing plays an important role in our paper. The
main differential equations 16 owe much to [1]. There are some important
differences. First we are using nonlinearity for the cell dynamics in contrast
to the Turing linear setting. Nonlinearity allows us to address issues of
stability, where the second derivative plays a crucial role and we are able to
use associated domains of the cell dynamics more in accord with the biology.
On the other hand Turing developed his work in a partial differential
equations framework, reaction diffusion equations, that reflect a continuum
perspective of the nature of cells. That leads to some applications in
morphogenesis, such as patterning in Zebra stripes [17, 18, 19, 20]. Our own
perspective differs. We feel that some of the basic features of morphogenesis
must deal with few cells (embryogenesis, cell differentiation).
\textbf{\medskip}

\noindent The recent work of Chua [11, 12] also develops Turing's
contributions in a different direction from our work. \textbf{\medskip}

\noindent Turing found an important example of the system of the same type we
used in Section 5. The example shows how a system that is stable without
diffusion becomes unstable in the presence of diffusion. \ Turing\ was
motivated to understand morphogenesis with this example of instability. The
example consists of two cells and two proteins. The variables$\ x_{1},$
$y_{1}$ represent concentrations of molecules (or proteins) for the first cell
and $x_{2},$ $y_{2}$ for the second cell. \ Turing's two cell
reaction-diffusion example can be written as:%

\begin{align}
\frac{dx_{1}}{dt} &  =\left(  5x_{1}-6y_{1}+1\right)  +0.5\left(  x_{2}%
-x_{1}\right) \tag{19}\\
\frac{dy_{1}}{dt} &  =\left(  6x_{1}-7y_{1}+1\right)  +4.5\left(  y_{2}%
-y_{1}\right) \nonumber\\
\frac{dx_{2}}{dt} &  =\left(  5x_{2}-6y_{2}+1\right)  +0.5\left(  x_{1}%
-x_{2}\right) \nonumber\\
\frac{dy_{2}}{dt} &  =\left(  6x_{2}-7y_{2}+1\right)  +4.5\left(  y_{1}%
-y_{2}\right)  ,\text{ \ \ \ }x_{1},y_{1},x_{2},y_{2}>0.\nonumber
\end{align}
The two cells are identical in this example and we can describe the cell
dynamics as $\frac{dx_{1}}{dt}=\left(  5x_{1}-6y_{1}+1\right)  $,
$\frac{dy_{1}}{dt}=\left(  6x_{1}-7y_{1}+1\right)  $. Is is easy to transform
System 19 into our form $\frac{d\mathbf{x}}{dt}\mathbf{=}$ $\left(
\mathbf{F-\mathbf{L}}\right)  \mathbf{x.}$\textbf{\medskip}

\noindent\textbf{Genome dynamics of the Turing example: }\medskip\ 

\noindent We now show that a key phenomenon of this example is the failure of
the monotonicity condition. That is necessary to give rise to instability
(morphogenesis).\medskip

\noindent Let us then study the monotonicity of Section 3 as well as a two
dimensional analysis for the Turing example. First we construct matrix
$\mathbf{A}$ for a single cell of the Turing example: $a=5,b=-7,c=6,d=7,$ and
$\mathbf{A}=\left(
\begin{array}
[c]{cc}%
5 & -6\\
6 & -7
\end{array}
\right)  .$ Here the Turing example assumes two identical cells and we can the
write monotonicity condition as $\left\langle \left(  5x-6y,\text{
}6x-7y\right)  \mathbf{,}\left(  x,\text{ }y\right)  \right\rangle <0$ for all
$x,y>0.$ Thus $5x^{2}-7y^{2}<0.$ If $x=2,y=1$ we can see that $5x^{2}%
-7y^{2}>0;$ therefore the Turing two cell example fails the monotonicity
condition (see the red dot in Figure 4). Monotonicity is only a sufficient
condition for stability. \ Now we check for stability.\medskip

\noindent The trace$\left(  \mathbf{A}\right)  $ $\mathbf{=}$ $-2,$ and
det$\left(  \mathbf{A}\right)  =1.$ Thus, the eigenvalues of $\mathbf{A}$ are
given by $\lambda_{i}(\mathbf{A})=-1,i=1,2$ (and the eigenvectors are given by
$v_{i}=\left(  1,1\right)  ,i=1,2).$ Therefore, the genome dynamics is stable
for one cell and hence for two cells.\medskip

\noindent\textbf{Diffusion dynamics}: \medskip

\noindent The diffusion dynamics in System 19 is expressed by the terms
$0.5\left(  x_{2}-x_{1}\right)  $ and $4.5\left(  y_{2}-y_{1}\right)  $.
\ Since the diffusion dynamics is represented by the negative laplacian matrix
as in Section 4, its eigenvalues are non postive. \medskip

\noindent\textbf{Full Dynamics:}\medskip

\noindent Combining genome dynamics and diffusion dynamics gives the Turing
example of System 19. The linear part of System 19 is%

\[
\mathbf{M}=\left(
\begin{array}
[c]{cccc}%
4.5 & -6 & 0.5 & 0\\
6 & -11.5 & 0 & 4.5\\
0.5 & 0 & 4.5 & -6\\
0 & 4.5 & 6 & -11.5
\end{array}
\right)  .
\]
System 19 has a unique equilibrium that is obtained by solving the right hand
side of the equation set equal to zero. Eigenvalues of $\mathbf{M}$ can be
computed to be $2.0,-1,-\allowbreak1,-14.$ Since there is a positive
eigenvalue, the system with diffusion is unstable. Therefore the Turing system
not only has a possibility of failure of stability but in fact it is
unstable.\medskip

\noindent Summarizing, the Jacobian at the equilibrium has four eigenvalues,
one of which is positive. Thus, the full system with diffusion is not stable.
\ In this way Turing showed that at the equilibrium the two cell example
without diffusion is stable, but with diffusion has lost stability. We have
remarked further on the role of monotonicity. \medskip

\noindent Smale [10] examined similar equations with nonlinear cell dynamics.
He considered each of two cells as having a global stable equilibrium, and
therefore the cells were "dead" in an abstract mathematical sense. But upon
coupling the two cells by diffusion, he proved that the resulting system has a
global periodic attractor, and hence the cells become "alive."\medskip

\noindent Towards this end Smale's work was a mathematical model similar to
the Turing two cell example but with dynamics of each cell not linear leading
to the model,
\begin{equation}
\frac{dx^{k}}{dt}=R\left(  x^{k}\right)  +\underset{i\in\text{ set of cells
neighboring }k{\mbox{\tiny \text{th}}}\text{ cell}}{\sum\mu_{ik}(x^{i}-x^{k}%
)}\tag{20}%
\end{equation}

\noindent where $k=1,...,N$ and $(x^{i}-x^{k})\in{\mathbb{R}}^{m}.$ The first
term above $R\left(  x^{k}\right)  $ gives the dynamics for the $k^{th}$ cell
and the second term describes the diffusion processes between cells. The
principal case considered by Smale is $m=4,$ $N=2,$ shows for the appropriate
choice of parameters $(R,\mu_{k}),$ the system has stable equilibria without
diffusion and with diffusion has a global periodic attractor. \ The Equation
20 is precisely a form of our main equations. Again the phenomenon depends on
the failure of monotonicity \bigskip

\noindent(\textbf{Easy) Conjecture 1}: Generically monotonicity of a linear
system on euclidean space is equivalent to all eigenvalues negative (and
real).\bigskip

\noindent\textbf{7. Lapse of emergence}\medskip

\noindent Here we discuss an avenue to study the departure from emergence
using our setting. Consider the Jacobian of $\frac{d{\mathbf{x}}}%
{dt}=\mathbf{F}{\mathbf{x}}-{\mathbf{L}}$ (Equation 16)\ at the equilibrium
${\mathbf{x}}_{0}$, that is
\[
\left(  \mathbf{D}\left(  \mathbf{\mathbf{F-L}}\right)  \right)
_{{\mathbf{x}}_{0}},\ \ \ {\mathbf{x}}_{0}=\left(  {\mathbf{x}}_{0,1}%
,...,{\mathbf{x}}_{0,m}\right)  ,
\]
where ${\mathbf{x}}_{0,i}$ is the equilibrium of the dynamics of the $i^{th}$
cell and ${\mathbf{x}}_{0,i}$ are all equal. The main cause of lapse of
emergence is the vanishing of determinant of $\left(  \mathbf{D}\left(
\mathbf{\mathbf{F-L}}\right)  \right)  _{{\mathbf{x}}_{0}}$. \ As in our paper
[20], the pitchfork bifurcation is signaled at a bifurcation parameter $\mu$,
where the det$(\mathbf{D}\left(  \mathbf{\mathbf{F-L}}\right)  )_{{\mathbf{x}%
}_{0}}$ first becomes zero. \ Since ${\mathbf{x}}_{0,i}$ are equal,
$\mathbf{L}{\mathbf{x}}_{0}=0$ and ${\mathbf{x}}_{0}$ belongs to the
$n-$dimensional harmonic space, kernel$\left(  \mathbf{L}\right)  .$
Generically $-\mathbf{L}$ is contracting to kernel$\left(  \mathbf{L}\right)
$, that is the eigenvalues of $-\mathbf{L}$ are $\lambda_{k}=0$ for $k\leq n$,
and for $k>n,\lambda_{k}<0.$\medskip

\noindent Now look at $\mathbf{D}\left(  \mathbf{\mathbf{F}}\right)
_{{\mathbf{x}}_{0}}$ $=\left(  \mathbf{D}\left(  \mathbf{\mathbf{F}}%
_{1}\right)  _{{\mathbf{x}}_{0,1}},\mathbf{D}\left(  \mathbf{\mathbf{F}}%
_{2}\right)  _{{\mathbf{x}}_{0,2}},...,\mathbf{D}\left(  \mathbf{\mathbf{F}%
}_{m}\right)  _{{\mathbf{x}}_{0,m}}\right)  $ on the basin $\mathbf{B}$. Each
$\mathbf{D}\left(  \mathbf{\mathbf{F}}_{i}\right)  _{{\mathbf{x}}_{0,i}}$ is
contracting before the bifurcation, say for $\mu<0.$ At $\mu=0$, one can
expect one of these contracting derivatives to become singular, for example
$\mathbf{D}\left(  \mathbf{\mathbf{F}}_{1}\right)  _{{\mathbf{x}}_{0,1}}$ with
one eigenvalue equal to zero and the rest are negative. This is the beginning
of the lapse of emergence in this scenario. Now restrict the dynamics to the
protein space of the first cell to study the bifurcation. \ In this protein
space we can expect the dynamics after the bifurcation to have two basins.
This is the setting of the pitchfork bifurcation paper [21]. This paper can be
used to examine the end of emergence in terms of cell division (symmetric or
asymmetric, or cancer) [22, 23].

\end{document}